*Cover page*

*Manuscript title:* DISARM: A Social Distributed Agent Reputation Model based on Defeasible Logic

*(Corresponding) First Author:*

**Kalliopi Kravari**

*Department of Informatics*

*Aristotle University of Thessaloniki*

*GR-54124, Thessaloniki, Greece*

*Email: kkravari AT csd.auth.gr*

*Tel: 00302310998231*

*Fax: 00302310998433*

*Second Author:*

**Nick Bassiliades**

*Department of Informatics*

*Aristotle University of Thessaloniki*

*GR-54124, Thessaloniki, Greece*

*Email: nbassili AT csd.auth.gr*

# DISARM: A Social Distributed Agent Reputation Model based on Defeasible Logic

**ABSTRACT** Intelligent Agents act in open and thus risky environments, hence making the appropriate decision about who to trust in order to interact with, could be a challenging process. As intelligent agents are gradually enriched with Semantic Web technology, acting on behalf of their users with limited or no human intervention, their ability to perform assigned tasks is scrutinized. Hence, trust and reputation models, based on interaction trust or witness reputation, have been proposed, yet they often presuppose the use of a centralized authority. Although such mechanisms are more popular, they are usually faced with skepticism, since users may question the trustworthiness and the robustness of a central authority. Distributed models, on the other hand, are more complex but they provide personalized estimations based on each agent's interests and preferences. To this end, this article proposes DISARM, a novel distributed reputation model. DISARM deals MASs as social networks, enabling agents to establish and maintain relationships, limiting the disadvantages of the common distributed approaches. Additionally, it is based on defeasible logic, modeling the way intelligent agents, like humans, draw reasonable conclusions from incomplete and possibly conflicting (thus inconclusive) information. Finally, we provide an evaluation that illustrates the usability of the proposed model.

# 1  Introduction

Intelligent Agents (IAs) act in open and thus risky environments, hence making the appropriate decision about the degree of trust that can be invested in a certain partner is vital yet really challenging [41]. Over the last few years, scientific research in this field has significantly increased. Most researchers tend to consider trust and reputation as key elements in the design and implementation of modern multi-agent systems (MASs). However, there is still no single, accepted definition of trust within the research community, although it is generally defined as the expectation of competence and willingness to perform a given task. Broadly speaking, trust has been defined in a number of ways in the literature, depending on the domain of use. Among these definitions, there is one that can be used as a reference point for understanding trust, provided by Dasgupta [12]. According to Dasgupta, trust is a belief an agent has that the other party will do what it says it will (being honest and reliable) or reciprocate (being reciprocative for the common good of both), given an opportunity to defect to get higher payoffs.

Trust, however, is much more than that; the uncertainties found in the modern MASs present a number of new challenges. More specifically, MASs are open distributed systems, sometimes large-scaled, which mean that the agents represent different stakeholders that are likely to be self-interested and might not always complete tasks requested from them. Moreover, given that the system is open, usually no central authority can control all the agents, which mean that agents can join and leave at any time. The problem is that this allows agents to change their identity and re-enter, avoiding punishment for any past wrong doing. One, more, risky feature of open systems is that when an agent first enters the system has no information about the other agents in that environment. Given this, the agent is likely to be faced with a large amount of possible partners with a different degree of efficiency and/or effectiveness.

To this end, a number of researchers were motivated by the understanding that some individuals, and thus their agents, may be dishonest, focusing eventually their efforts on agents' reputation. In general, reputation is the opinion of the public towards an agent. Reputation allows agents to build trust, or the degree to which one agent has confidence in another agent, helping them to establish relationships that achieve mutual benefits. Hence, reputation (trust) models help agents to decide who to trust, encouraging trustworthy behavior and deterring dishonest participation by providing the mean through which reputation and ultimately trust can be quantified [43]. Hence, as intelligent agents are gradually enriched with Semantic Web technology [18, 6, 5], acting on behalf of their users with limited or no human intervention, their ability to perform assigned tasks is scrutinized. To this end, plenty of trust and reputation models have been proposed in different perspectives, yet they often presuppose the use of a centralized authority [41]. Although such reputation mechanisms are more popular, they are usually faced with skepticism, since in open MASs agents represent different owners, who may question the trustworthiness and the robustness of a central authority.

On the other hand, distributed reputation models are typically more complex and require a lot of communication in order agents to exchange their ratings. These models have no centralized system manager; hence each agent has to overcome the difficulty of locating ratings and develop somehow a subjective estimation by itself using its own resources. No global or public reputation exists. The reputation built in this way is thus personalized and sometimes difficult to reach. However, a distributed reputation system is more flexible in building agents' reputation, since it is quite easy for an agent to develop differentiated trust in other agents based on its interests and purposes. Yet, beyond the traditional choice of centralized or distributed approach, there is an even more challenging decision; what should be taken into account in order to estimate the reputation of an agent, interaction trust or witness reputation [2, 32]. In other words, an agent's direct experience or reports provided by others, respectively.

Broadly speaking, both approaches have limitations. For instance, if the reputation estimation is based only on direct experience, it would require a long time for an agent to reach a satisfying estimation level. This is because, when an agent enters an environment for the first time, it has no history of interactions with the other agents in the environment. Thus, it needs a long time to reach a sufficient amount of interactions that could lead to sufficient information. On the other hand, models based only on witness reports could not guarantee reliable estimation as self-interested agents could be unwilling or unable to sacrifice their resources in order to provide reports. Hence, models based only on one or the other approach typically cannot guarantee stable and reliable estimations. To this end, in order to overcome these drawbacks, a number of hybrid models that combine both interaction trust and witness reputation were proposed [2, 20, 21, 32]. However, most of hybrid models either have fixed proportion of their active participation in the final estimation or leave the choice to the final user. Although these approaches have significant advantages, sometimes they may lead to misleading estimations. Users may have little or no experience and thus take wrong decisions that could lead to wrong assessments, whereas fixed values provide just generic estimations. Our goal is not to estrange the users from the decision making process, but to help them, and their agents, to make better decisions.

To this end, this article proposes a novel distributed reputation model, called DISARM, that combines both interaction trust and witness reputation. DISARM is a knowledge-based approach, based on well-established estimation parameters [8, 9, 16, 19, 22, 49, 48], that provides a more intuitive method for non-technical users. More specifically, its intention is to reduce the disadvantages of the common distributed hybrid approaches, such as the difficulty in locating ratings, and provide a mechanism for modeling the way intelligent agents, like humans, draw reasonable conclusions from incomplete and possibly conflicting (thus inconclusive) information. This is achieved by designing and implementing a reputation mechanism based on social principles and defeasible logic. Concerning the social aspect of the model, DISARM proposes an approach where agents are enabled to establish, through their interactions, and maintain relationships, much as individual do in real life. More specifically, DISARM considers agents acting in the environment as a social network which determines the proximity relationships among them. In this context, all known agents create a network, which is expanded whenever new agent interactions take place in the environment. The advantage of this approach is that it allows agents to communicate with previously known and well-rated agents, locating, quite fast, ratings with small bandwidth cost.

Concerning the modeling mechanism, although it is logic-independent, DISARM proposes the use of defeasible logic, a logic that has the notion of rules that can be defeated, allowing an existing belief to turn false, making it nonmonotonic [35, 39]. In a fundamental sense, nonmonotonic logics occupy undoubtedly prominent position among the disciplines investigating intelligent reasoning about complex and dynamic situations. Thus, permitting agents to arrive at defeasible conclusions, leads to more realistic assessments similar to human reasoning. Additionally, defeasible logic is part of a more general area of research, defeasible reasoning, which is notable for its low computational complexity [31].

Moreover, we provide an evaluation that illustrates the usability of the proposed model. The rest of the article is organized as follows. In Section 2, we present a brief overview of defeasible logic. Section 3 presents DISARM and its contribution. In Section 4, DISARM's evaluation is presented, demonstrating the added value of the approach. Section 5 discusses related work, and Section 6 concludes with final remarks and directions for future work.

## 2   Defeasible logic

Defeasible logic (DL), introduced by Nute [34, 36] with a particular concern about efficiency and implementation, is part of a more general area of research, defeasible reasoning [39, 38]. Over the years the logic has been developed and extended while several variants have been proposed. Yet, DL remains a simple and efficient rule based nonmonotonic formalism that deals with incomplete and conflicting information. More specifically, DL has the notion of rules that can be defeated; hence it derives plausible conclusions from partial and sometimes conflicting information. These conclusions, despite being supported by the currently available information, could nonetheless be rejected in the light of new, or more refined, information.

Compared to other more mainstream nonmonotonic approaches, e.g. [42, 13], this approach offers among others enhanced representational capabilities and low computational complexity. Moreover, DL in contrast with traditional deductive logic, allows the addition of further propositions to make an existing belief false, making it nonmonotonic [26]. In a fundamental sense, nonmonotonic logics occupy undoubtedly prominent position among the disciplines investigating intelligent reasoning about complex and dynamic situations / environments. Hence, one of the main interests in DL is in the area of agents [14]. DL, being a nonmonotonic logic, is capable of modeling the way intelligent agents, like humans, draw reasonable conclusions from inconclusive information, leading to more realistic conclusions and assessments similar to human reasoning.

Knowledge in DL is represented in terms of facts, rules and superiority relations. Facts are indisputable statements, represented either in form of states of affairs (literal and modal literal) or actions that have been performed. Rules describe the relationship between a set of literals (premises) and a literal (conclusion). Rules are divided into strict rules, defeasible rules and defeaters. Strict rules are rules in the classical sense, whenever the premises are indisputable, e.g. facts, then so is the conclusion. Thus, they can be used for definitional clauses. Defeasible rules, on the other hand, are rules that can be defeated by contrary evidence. Defeaters are rules that cannot be used to draw any conclusions. Their only use is to prevent some conclusions. The form of rules, in symbolic and d-POSL syntax [25], are presented in Table 1. Finally, the superiority relation is a binary relation defined over the set of rules, which determines the relative strength of two (conflicting) rules, i.e. rules that infer conflicting literals.

**Table 1.** Rules in Defeasible Logic.

| RULE TYPE | RULE FORM | D-POSL SYNTAX |
|---|---|---|
| STRICT RULES | A1, … , An → B | B:-A1, … , An |
| DEFEASIBLE RULES | A1, … , An => B | B:=A1, … , An |
| DEFEATERS | A1, … , An ~> B | B:~A1, … , An |

The main concept in DL is that it does not support contradictory conclusions, but it tries to resolve conflicts. Hence, in cases where there is some support for concluding A, but there is also support for concluding ¬A, no conclusion is derived unless one of the two rules that support these conflicting conclusions has priority over the other. This priority is expressed through a superiority relation among rules which defines priorities among them, namely where one rule may override the conclusion of another rule. Yet, conclusions can be classified as definite or defeasible. A definite conclusion is a conclusion that cannot be withdrawn when new information is available. A defeasible conclusion, on the other hand, is a tentative conclusion that might be withdrawn in the future. In addition, the logic is able to tell whether a conclusion is or is not provable, hence there are four possible types of conclusions; positive definite, negative definite, positive defeasible and negative defeasible.

Positive definite conclusions are provable using only facts and strict rules while negative definite conclusions are not provable by using these. Positive defeasible conclusions can be defeasible proved, while negative defeasible conclusions are not even defeasibly provable. Strict derivations are obtained by forward chaining of strict rules, while a defeasible conclusion A can be derived if there is a rule whose conclusion is A, whose premises have either already been proved or given in the form of facts and any stronger rule whose conclusion is ¬A (the negation of A) has premises that fail to be derived or the latter rule has been defeated by an even stronger rule. In this context, a special case of conflict is between different positive literals, all derived by different defeasible rules, whereas only one should be derived. "Conflicting literals" are defined through a conflict set and the conflict is resolved through superiorities. [7]

## 3   DISARM

The proposed model is called DISARM and it is a distributed, hybrid, rule-based reputation model. DISARM uses defeasible logic in order to combine in a practical way all available ratings, both those based on the agent's personal experience and those provided by known and/or unknown third parties. This model aims not only at reducing the disadvantages of the common distributed approaches, such as the difficulty in locating ratings, but mainly it aims at improving the performance of the hybrid approach by providing an intuitive decision making mechanism. DISARM aims at providing a distributed mechanism based on defeasible logic that would be able to model the way humans think, infer and decide.

## 3.1 Main principles of the DISARM Model

For purposes of better understanding, we present here the main principles of our model. First of all, DISARM has no centralized authority since it is a distributed model. Hence, it is each agent's responsibility to locate ratings and use the model. In this context, even if more than one agents have available the same ratings, they probably will come out with different estimations and as a result they will take different decisions.

Time, evolution over time in particular, is an important issue since it reflects the behavior of an agent. More specifically, in dynamic environments such as MASs, agents may change their objectives at any time. For instance, a typical dishonest agent could provide quality services over a period to gain a high reputation score, and then, profiting from that high score could provide low quality services. Hence, time should be and it is taken into account in the proposed model. Yet, DISARM allows agents to decide on their own about what they consider important. To this end, it is up to each agent's strategy to determine the value of time. Agents could take into account all the available ratings or only the latest; e.g. those referred to last week, last month or last year.

Taking into account the latest ratings leads undoubtedly to an up-to-date overview, however it could be misleading. For instance, in this limited time period, a typical dishonest agent could temporary improve its behavior or, on the other hand, a typical reliable agent, facing a problem, could temporary act faulty, transformed into a mercenary and malicious agent. Hence, it is a risk to take into account only part of the available ratings, although there is sometimes significant gain in time and computational cost. In contrary, taking into account all available ratings leads to an overview of an agent's behavior history but it costs in terms of storage space, execution time and computational power.

DISARM, however, is a distributed model which means that locating ratings is a quite challenging process. The rating records could be always there but usually they are unreachable since various agents may join or leave the system at any time. For instance, sometimes only a few ratings are available; e.g. personal experience could be missing and/or appropriate witnesses could be difficult to locate. On the other hand, sometimes there is a large amount of available ratings but taking all of them into account has significant computational cost. Moreover, these ratings may significantly differ. In this context, DISARM integrates an indication of how likely is the assessment to be proved correct based on the variability of ratings that were taken into account. In other words, DISARM allows agents to be informed about the possibility of wrong estimation and loss.

Another important issue that DISARM deals with is the trust relationships that agents build and maintain over time, much as individuals do in real world. For instance, if an agent is satisfied with a partner, probably it will prefer to interact again with that partner in the future. On the other hand, if it is disappointed by a partner, it will avoid interacting again with that partner. To this end, DISARM proposes the use of two lists, called white-list and black-list. Each agent stores in its white-list the names of its favored partners while in its black-list it stores those that should be avoided. The decision about who will be added in each list is taken by the agent itself. More specifically, each agent is equipped with a rule-based decision-making logic which enables it to decide upon its partners, adding them, if necessary, to the appropriate list. Hence, it will be easy for the agent to locate a well-known old partner that will do the job and at the same time avoid a fraud. Moreover, a user is much more likely to believe statements from a trusted acquaintance than from a previously known dishonest agent or a stranger.

Finally, additionally to the difficulty to locate ratings is the difficulty to locate really useful ratings. For instance, sometimes agents are involved in important and crucial for them interactions whereas sometimes they are involved in simple interactions of minor importance. Hence, the question is which of them should be taken into account in order to get a representative estimation. To this end, DISARM adopts the use of two more parameters for each rating; namely importance and confidence. Importance indicates how critical the transaction was for the rating agent while confidence gives an estimation of the agent's certainty for that rating.

## 3.2 Rating parameters

Taking into account the proper parameters for an assessment is a really challenging task. They should be carefully chosen in order to reflect the agents' abilities. Besides, an efficient decision making mechanism has to rely on carefully selected data and a straightforward and efficient rating procedure. Although, a thorough overview of related literature is out of the scope of this article, we tried to catch out parameters, or factors for others, that are usually referred either explicitly or implicitly in reputation models and metrics, e.g. [8, 9, 16, 19, 22, 49, 48]. To this end, DISARM uses for its needs six properties; namely response time, validity, completeness, correctness, cooperation and outcome feeling.

*Response time* refers to the time that an agent needs in order to complete the tasks that it is responsible for. Time is the only parameter that is always taken into account in the literature. *Validity* describes the degree that an agent is sincere and credible. An agent is sincere when it believes what it says, whereas it is credible when what it believes is true in the world. Hence, an agent is valid if it is both. Validity is not always such called, yet in most cases there are parameters that attempt to indicate how sincere and/or credible an agent is. *Completeness*, on the other hand, describes the degree that an agent says what it believes while what it believes is true in the world. In other words, completeness is the inverse of validity, indicating how honest and realistic an agent is. Completeness is usually implicitly referred, as an attempt to rate dishonest and fraud behavior.

Moreover, *correctness* refers to an agent's providing services. An agent is correct if its provided service is correct with respect to a specification. Correctness, no matter how it is called, is, actually, the second most used parameter after time. *Cooperation* is the willingness of an agent who is being helpful by doing what is wanted or asked for. Cooperation is not, usually, handled as separate parameter, however, it is an important task in distributed social environments, such as MASs. Finally, the *outcome feeling* is a general feeling of satisfaction or dissatisfaction related to the transaction outcome; namely it indicates if the transaction was easy and pleasant with a satisfying result or not. Usually, it is referred as the degree of request fulfillment.

However, although these six parameters are, usually, taken into account in one way or another, they are not necessarily binding. Some of them could be replaced by other more domain-specific parameters depending on the domain of use, e.g. agents acting in E-Commerce transactions. Yet, our intention, here, in the context of DISARM, is to provide general purpose parameters that will be able to reflect the common critical characteristics of each agent in the community. In other words, DISARM takes into account parameters that can provide an overview of each agent's behavior. Hence, consider an agent A establishing an interaction with an agent X; agent A can evaluate the other agent's performance and thus affect its reputation. The evaluating agent (A) is called *truster* whereas the evaluated agent (X) is called *trustee*. Of course, for some interactions an agent can be both truster and trustee, since it can evaluate its partner while it is evaluated by that partner at the same time. After each interaction in the environment, the truster has to evaluate the abilities of the trustee in terms of response time, validity, completeness, correctness, cooperation and outcome feeling. DISARM, however, is a distributed model hence truster does not have to report its ratings but just to save them for future use.

Yet, in order to remember how important was the transaction for it and how confident it was for its rating, the agent has to associate two more values to the rating, as was already discussed, namely confidence and importance. Additionally, since time is considered an important aspect in DISARM's decision making process, each rating is associated with a time stamp (t), e.g. in the form YYMMDDHHMMSS or YYMMDD or HHMMSS or it can be represented even as an integer in case of experimental simulations, indicating the transaction's time point. Hence, taking the above into account the truster's rating value (r) in DISARM is a tuple with eleven elements: *(truster, trustee, t, response time, validity, completeness, correctness, cooperation, outcome feeling, confidence, transaction value)*. Notice, that although each truster agent stores its own ratings, we include the variable truster in the rating value. This is because the truster may forward its ratings to other agents; hence, these agents should be able to identify the rating agent for each rating they receive. In DISARM, the rating values vary from 0.1 (terrible) to 10 (perfect); $r \in [0.1, 10]$, except confidence and transaction values that vary from 0 (0%) to 1 (100%).

### 3.3 Rule-based decision mechanism

Defining the rating values is the first step towards an efficient reputation model, the core of the approach, however, is its decision making mechanism. The distributed reputation models have invariably to deal with a range of complex issues related to the decision making process, such as locating ratings. Hence, DISARM aims at providing a trust estimation procedure much as individuals do in real world, where they build and maintain trust relationships over time. To this end, DISARM simulates their decision making process, proposing a set of strict and defeasible rules, in a practical, intuitive approach.

### 3.3.1 Rating procedure

First of all, as soon as, an interaction ends each agent evaluates its partner in terms of response time, validity, completeness, correctness, cooperation and outcome feeling. Then it adds its confidence and a value indicating the importance of the transaction (transaction value). When all values are got together, the rating agent (truster) adds its name, the trustee's name and the current time point (t), forming the final rating value (r) as a tuple with eleven elements. This tuple is presented below in the compact d-POSL syntax [25] of defeasible RuleML [3]. A syntax that will be used throughout this article in order to express in a compact way the data (ratings) and rules (strict and defeasible rules used in the decision making process) of our approach. To this end, the truster's rating (r) is the fact:

$rating(id \rightarrow rating's\_id, truster \rightarrow truster's\_name, trustee \rightarrow trustee's\_name, t \rightarrow time\_stamp,$

$response\_time \rightarrow value_1, validity \rightarrow value_2, completeness \rightarrow value_3,$

$correctness \rightarrow value_4, cooperation \rightarrow value_5, outcome\_feeling \rightarrow value_6,$

$confidence \rightarrow value_7, transaction\_value \rightarrow value_8).$

Additionally, an example rating provided by agent (A) truster for the agent (X) trustee could be:

$rating(id \rightarrow 1, truster \rightarrow A, trustee \rightarrow X, t \rightarrow 140630105632, response\_time \rightarrow 9, validity \rightarrow 7,$

$completeness \rightarrow 6, correctness \rightarrow 6, cooperation \rightarrow 8,$

$outcome\_feeling \rightarrow 7, confidence \rightarrow 0.9, transaction\_value \rightarrow 0.8).$

Next, truster stores this rating to its repository. However, as already mentioned, agents compliant with DISARM use two lists, additionally to their rating repository; white-list and black-list. More specifically, these lists are two separate repositories, one for storing promising partners (white-list) and one for those partners that should be avoided (black-list). Hence, truster has also to decide whether it should add the trustee to its white (or black) list or not. Obviously, the decision is based on what it is considered as a promising (or terrible on the other hand) partner. Promising is a partner if it acts responsibly and it provides high quality services or products. A partner is responsible if it is cooperative, responds fast and leaves a positive feeling at the end of the transaction.

Of course, each agent has a different degree of tolerance, thus, what may be fast for an agent could be slow for another. Hence, each agent has some thresholds that determine the lowest accepted value for each parameter; namely response time, validity, completeness, correctness, cooperation and outcome feeling. Moreover, from each agent's perspective a parameter could be more important than others. For instance, an agent could consider response time the most important aspect, perhaps not the only, in deciding whether its partner could be characterized good or bad. Hence, truster classifies its partner in relation to a reason, e.g. response time. In this context, rule $r_1$, presented below, indicates that if all values are higher than the truster's associate thresholds then the trustee's behavior is considered good.

$r_1$: *good_behavior(time → ?t, truster→ ?a, trustee→ ?x, reason → response_time)* :-

*response_time_threshold(?resp), validity_threshold(?val),*

*completeness_threshold(?com), correctness_threshold(?cor),*

*cooperation_threshold(?coop), outcome_feeling_threshold(?outf),*

*rating(id→?idx, time → ?t, truster→ ?a, trustee→ ?x,*

*response_time→?respx, validity→?valx, completeness→?comx,*

*correctness→?corx, cooperation→?coopx, outcome_feeling→?outfx),*

*?respx>?resp, ?valx>?val, ?comx>?com, ?corx>?cor,*

*?coopx>?coop, ?outfx>?outf.*

Where response_time (reason) could be replaced by one of the rest parameters, namely validity, completeness, correctness, cooperation and outcome_feeling. On the other hand, trustee's behavior is consider disappointing and, thus, bad in relation to a reason/parameter, if trustee's rate for this parameter is lower than the truster's thresholds. Rules $r_2$ to $r_7$ present the group of rules that characterize the behavior of an agent as bad depending on a specific reason.

$r_2$: *bad_behavior(time → ?t, truster→ ?a, trustee→ ?x, reason → response_time) :-*

    *rating(id→?idx, time → ?t, truster→ ?a, trustee→ ?x, response_time→?respx),*

    *response_time_threshold(?resp),*

    *?respx<=?resp.*

$r_3$: *bad_behavior(time → ?t, truster→ ?a, trustee→ ?x, reason → validity):-*

    *rating(id→?idx, time → ?t, truster→ ?a, trustee→ ?x, validity→?valx),*

    *validity_threshold(?val),*

    *?valx<=?val.*

$r_4$: *bad_behavior(time → ?t, truster→ ?a, trustee→ ?x, reason→ completeness) :-*

    *rating(id→?idx, time → ?t, truster→ ?a, trustee→ ?x, completeness→?comx),*

    *completeness_threshold(?com),*

    *?comx<=?com.*

$r_5$: *bad_behavior(time → ?t, truster→ ?a, trustee→ ?x, reason→ correctness):-*

    *rating(id→?idx, time → ?t, truster→ ?a, trustee→ ?x, correctness→? corx),*

    *correctness_threshold(?cor),*

    *?corx<=?cor.*

$r_6$: *bad_behavior(time → ?t, truster→ ?a, trustee→ ?x, reason→ cooperation) :-*

    *rating(id→?idx, time → ?t, truster→ ?a, trustee→ ?x, cooperation→?coopx),*

    *cooperation_threshold(?coop),*

    *?coopx<=?coop.*

$r_7$: *bad_behavior(time → ?t, truster→ ?a, trustee→ ?x, reason → outcome_feeling):-*

$rating(id \rightarrow ?idx, time \rightarrow ?t, truster \rightarrow ?a, trustee \rightarrow ?x, outcome\_feeling \rightarrow ?outfx)$,

$outcome\_feeling\_threshold(?outf)$,

$?outfx <= ?outf$.

However, characterizing a trustee's behavior good or bad does not, necessary, mean that this trustee will be added to the truster's white or black list, respectively. This decision is left to the truster's private strategy and it could vary greatly from agent to agent. For instance, a truster could be lenient and, thus, it might add quite easily trustees to its white-list. Another truster might expect to see good behavior several times either for the same reason ($r_8$, where *?self* represents the truster itself) or for a number of reasons ($r_9$), before adding a trustee to its white-list. Similarly, a truster might expect to face a trustee's bad behavior more than one times either for the same reason ($r_{10}$) or for a number of reasons ($r_{11}$), before adding the trustee to its black-list. Hence, a strict truster would easily add trustees to its black-list but not to its white-list whereas a lenient would give more changes before adding a trustee to its own black-list.

$r_8$: $add\_whitelist(trustee \rightarrow ?x, time \rightarrow ?t3) :=$

$good\_behavior(time \rightarrow ?t1, truster \rightarrow ?self, trustee \rightarrow ?x, reason \rightarrow ?r)$,

$good\_behavior(time \rightarrow ?t2, truster \rightarrow ?self, trustee \rightarrow ?x, reason \rightarrow ?r)$,

$good\_behavior(time \rightarrow ?t3, truster \rightarrow ?self, trustee \rightarrow ?x, reason \rightarrow ?r)$,

$?t2 > ?t1, ?t3 > ?t2$.

$r_9$: $add\_whitelist(trustee \rightarrow ?x, time \rightarrow ?t3) :=$

$good\_behavior(time \rightarrow ?t1, truster \rightarrow ?self, trustee \rightarrow ?x, reason \rightarrow ?r1)$,

$good\_behavior(time \rightarrow ?t2, truster \rightarrow ?self, trustee \rightarrow ?x, reason \rightarrow ?r2)$,

$good\_behavior(time \rightarrow ?t3, truster \rightarrow ?self, trustee \rightarrow ?x, reason \rightarrow ?r3)$,

$?t2 > ?t1, ?t3 > ?t2$.

$r_{10}$: $add\_blacklist(trustee \rightarrow ?x, time \rightarrow ?t2) :=$

$\quad$ bad_behavior(time → ?t1, truster→ ?self, trustee→ ?x, reason → ?r),

   $\quad$ bad_behavior(time → ?t2, truster→ ?self, trustee→ ?x, reason → ?r),

   $\quad$ ?t2 > ?t1.

$r_{11}$: add_blacklist(trustee→ ?x, time → ?t3) :=

   $\quad$ bad_behavior(time → ?t1, truster→ ?self, trustee→ ?x, reason → ?r1),

   $\quad$ bad_behavior(time → ?t2, truster→ ?self, trustee→ ?x, reason → ?r2),

   $\quad$ bad_behavior(time → ?t3, truster→ ?self, trustee→ ?x, reason → ?r3),

   $\quad$ ?t2 > ?t1, ?t3 > ?t2, ?r2 ≠ ?r1, ?r3 ≠ ?r2, ?r3 ≠ ?r1.

Mention that the above rules are defeasible since they are part of the truster's preferences (private strategy). The priority relationship among them could vary from case to case and it is left to the truster. Other theories could, also, be used depending on the requirements and preferences a truster, its user in particular, has. Next, as soon as, the truster decides upon who should be added to the white-list and/or the black-list, it proceeds to the next part of rules ($r_{12}$ - $r_{15}$) where the addition is, actually, carried out.

$r_{12}$: blacklist(trustee→ ?x, time → ?t) :=

   $\quad$ ¬whitelist(trustee→ ?x, time → ?t1),

   $\quad$ add_blacklist(trustee→ ?x, time → ?t2),

   $\quad$ ?t2 > ?t1.

$r_{13}$: ¬blacklist(trustee→ ?x, time → ?t2) :=

   $\quad$ blacklist(trustee→ ?x, time → ?t1),

   $\quad$ add_whitelist(trustee→ ?x, time → ?t2),

   $\quad$ ?t2 > ?t1.

$r_{14}$: whitelist(trustee→ ?x, time → ?t) :=

   $\quad$ ¬blacklist(trustee→ ?x, time → ?t1),

add_whitelist(trustee→ ?x, time → ?t2),

?t2 > ?t1.

$r_{15}$: ¬whitelist(trustee→ ?x, time → ?t2) :=

whitelist(trustee→ ?x, time → ?t1),

add_blacklist(trustee→ ?x, time → ?t2),

?t2 > ?t1.

As a result, the truster's white-list, $WL_A \equiv \{X_i, ... X_n\}$, finally, includes the names ($?x \rightarrow X_i$) of all its favored agents ($r_{16}$ - $r_{17}$) whereas its black-list, $BL_A \equiv \{X_j, ... X_m\}$, includes the agents that it would prefer to avoid ($r_{18}$ - $r_{19}$).

$r_{16}$: WL(trustee→ ?x) :=

whitelist(trustee→ ?x, time → ?t1),

not(¬whitelist(trustee→ ?x, time → ?t2), ?t2 > ?t1)).

$r_{17}$: ¬WL(trustee→ ?x) :~

¬whitelist(trustee→ ?x, time → ?t1),

not(whitelist(trustee→ ?x, time → ?t2), ?t2 > ?t1)).

$r_{18}$: BL(trustee→ ?x) :=

blacklist(trustee→ ?x, time → ?t1),

not(¬blacklist(trustee→ ?x, time → ?t2), ?t2 > ?t1)).

$r_{19}$: ¬BL(trustee→ ?x) :~

¬blacklist(trustee→ ?x, time → ?t1),

not(blacklist(trustee→ ?x, time → ?t2), ?t2 > ?t1)).

### 3.3.2 Locating ratings

A major challenge for open distributed and sometimes large-scaled (multi-agent) systems is how to locate ratings among the rest of the community. The simplest and most common approach in such a distributed environment is to send a request message [50, 23]. Yet, the question is how and to whom this message should be sent directly and probably propagated by the direct and indirect receivers. To this end, using as a guide research on peer-to-peer networks [1], there are two core ways to propagate messages in order to locate peers (or ratings in our case) [33]. The first approach assigns a maximum time-to-live (TTL) parameter to each request message hence the requesting peer sends the message to its neighbors, who relay it to their own neighbors and so on until the time-to-live value is reached. The second approach allows peers to relay the message only to one neighbor at time, since they have to wait the response from a neighbor before forward the message to another neighbor. The first approach increases the communication cost, leading to significant higher bandwidth consumption but partners (and so ratings) are located fast. On the other hand, the second approach requires low bandwidth but it leads to time delays since more time is required to get feedback for the requests.

Over the last years, a number of researchers have proposed approaches that try to reduce bandwidth or improve response time (e.g. [30, 40]), mainly focusing on how to reach good and far away peers. Although, it is out of the scope of this article to research or improve peer-to-peer message propagate protocols, we were inspired by these approaches. To this end, DISARM, proposes a more intuitive approach where agents take advantage of their previously established relationships in order to propagate their new requests, finding, quite fast, ratings with small bandwidth cost. More specifically, although the notion of neighbors does not exist in MASs, agents can use previously known partners in a similar point of view. To this end, in DISARM

MASs are considered as social networks of agents. Such a social network can actually be represented by a social graph; a graph based on previously known agents either good (white-list) or bad (black-list). Hence, the known agents of an agent are, in our point of view, its neighbors. Using the knowledge represented by the social graph, DISARM is able to determine the proximity relationships among agents in the environment. In this context, it is easier for an agent to propagate its requests and eventually locate appropriate ratings.

Hence, an agent A that wants to collect ratings referred to agent X, does not send a request message to all agents but only to those stored in its white-list. The motivation behind this action is the fact that a user is much more likely to believe statements from a trusted acquaintance than from a previously known dishonest agent or a stranger. Yet, these previously known and well behaved agents may have no interaction history with agent X. This could lead to limited or zero feedback for the requesting agent A. To this end, adopting the notion of TTL, in DISARM each ratings request message is accompanied with a TTL value, where TTL represents the hops in the graph. Hence, each request is characterized by its horizon (TTL value) that determines how far the message will be propagated in the environment. In other words, the requesting agent determines if it is allowed (TTL $\neq$ 0) for its known (white-listed agents) to ask their own known agents, namely agents included in their white lists ($WL \equiv \{X_k, ... X_l\}$) and so on. Hence, the request message will be propagated in steps; each time an agent receives such a request forwards it to its well-behaved known agents, if it is allowed (TTL $\neq$ 0), reducing the TTL value by one. However, if the requesting agent is included in the black-list then its request message is ignored.

Moreover, the TTL value acts as a termination condition so that messages are not propagated indefinitely in the MAS; whenever an agent receives a request message with zero TTL does not forward the message. Finally, each agent will return, following the reverse path of the request, both its ratings and those provided by its partners, which eventually will be received by the initial requesting agent A. The above rule-based framework is, actually, logic independent since it can be implement in any logic. Yet, in DISARM, we use defeasible logic, as already mentioned, for purposes of simplicity and efficiency.

Rules $r_{20}$-$r_{24}$, below, model the above mentioned behavior. More specifically, rule $r_{20}$ initiates the ratings requests by sending it to all agents ?r in the white list, along with the TTL parameter. Rule $r_{21}$ is responsible for answering back to the requesting agent about the requested agent's rating is such a previous experience exists in the local knowledge base. Rule $r_{22}$ is responsible for forwarding a received request to agents in the white list if the TTL is still positive, by decreasing it at the same time. Rule $r_{23}$ is a defeater rule that defeats rules $r_{22}$, namely it will block answering back to bad agents. Finally, rule $r_{24}$ will store in the local knowledge base received ratings, is the sender is not in the blacklist. Notice that in this rule we use negation as failure, meaning that if *BL*(?s) fails during execution then not(*BL*(?s)) will succeed in order to determine if a sender agent does not belong to the blacklist.

$r_{20}$: send_message(sender→?self, receiver→?r, msg →*request_reputation*(about→?x,ttl→?t)) :=
    ttl_limit(?t), WL(?r), locate_ratings(*about*→?x).

$r_{21}$: send_message(sender→?self, receiver→?s,
    msg →*rating*(id→*rating's_id*, truster→ ?self, trustee→ ?x, t→*time_stamp*,
    *response_time*→*value₁*, *validity*→*value₂*, *completeness*→*value₃*,
    *correctness*→*value₄*, *cooperation*→*value₅*, *outcome_feeling*→*value₆*,
    *confidence*→*value₇*, *transaction_value*→*value₈*)) :=
    receive_message(sender→?s, receiver→?self, msg →*request_rating*(about→?x,ttl→?t)),
    rating(id→*rating's_id*, truster→ ?self, trustee→ ?x, t→*time_stamp*,
    *response_time*→*value₁*, *validity*→*value₂*, *completeness*→*value₃*,
    *correctness*→*value₄*, *cooperation*→*value₅*, *outcome_feeling*→*value₆*,

$$confidence \rightarrow value_7, transaction\_value \rightarrow value_8).$$

$r_{22}$: send_message(sender→?s, receiver→?r, msg →*request_reputation(about→?x,ttl→?t1)*):=

    receive_message(sender→?s, receiver→*?self*,

        msg →*request_rating(about→?x,ttl→?t)*),

    ?t >0, WL(?r), ?t1 is ?t - 1.

$r_{23}$: ¬send_message(sender→?self, receiver→?s, msg →?m) :~

    send_message(sender→?self, receiver→?s, msg →?m),

    BL(?s).

$r_{24}$: rating(id→*rating's_id*, truster→ ?x, trustee→ ?y, t→*time_stamp*,

      *response_time→value$_1$, validity→value$_2$, completeness→value$_3$*,

      *correctness→value$_4$, cooperation→value$_5$, outcome_feeling→value$_6$*,

      *confidence→value$_7$, transaction_value→value$_8$*) :=

    receive_message(sender→?s, receiver→*?self*,

      msg →rating(id→*rating's_id*, truster→ ?x, trustee→ ?y, t→*time_stamp*,

      *response_time→value$_1$, validity→value$_2$, completeness→value$_3$*,

      *correctness→value$_4$, cooperation→value$_5$, outcome_feeling→value$_6$*,

      *confidence→value$_7$, transaction_value→value$_8$*)),

    not(BL(?s)).

### 3.3.3 Discarding ratings

As soon as, all available ratings are collected, an important decision has to be made; which ratings will be taken into account. Ratings represent the experience of the involved parties, which is distinguished to direct (agent's direct experience $PR_X$) and indirect experience. Indirect experience is divided in two categories, ratings provided by strangers ($SR_X$) and reports provided by known agents due to previous interactions. In this context, $r_{25}$ determines which agents are considered as known. Additionally to that, known agents are divided to three more categories; agents included in the WL white-list ($WR_X$), agents included in the BL black-list ($BR_X$) and the rest known agents ($KR_X$). It is well known that using different opinions of a large group maximizes the possibility of crossing out unfair ratings. Hence, using both direct and indirect experience could lead to more truthful estimations.

$r_{25}$:    *known(agent→?x)* :-

   *rating(id→?id$_x$, truster→?self, trustee→?x)*.

However, sometimes one or more rating categories are missing, for instance, a newcomer has no personal experience and, thus, there are no available ratings ($PR_X$). To this end, we wish to ground our conclusions in trust relationships that have been built and maintained over time, much as individuals do in real world. For instance, a user is much more likely to believe statements from a trusted acquaintance than from a stranger. Thus, personal opinion (PR) is more valuable than acquaintances opinion (KR), which in turn is more valuable than strangers' opinion (SR).Furthermore, previously known and black-listed agents are generally considered unreliable than trusted agents (known agents or agents in the white-list) and, thus, they are ignored. Finally, agents in the white-list (WR) are usually more trusted than mere acquaintances (KR). In this context, the relationship among the rating categories is presented graphically in Fig. 1.

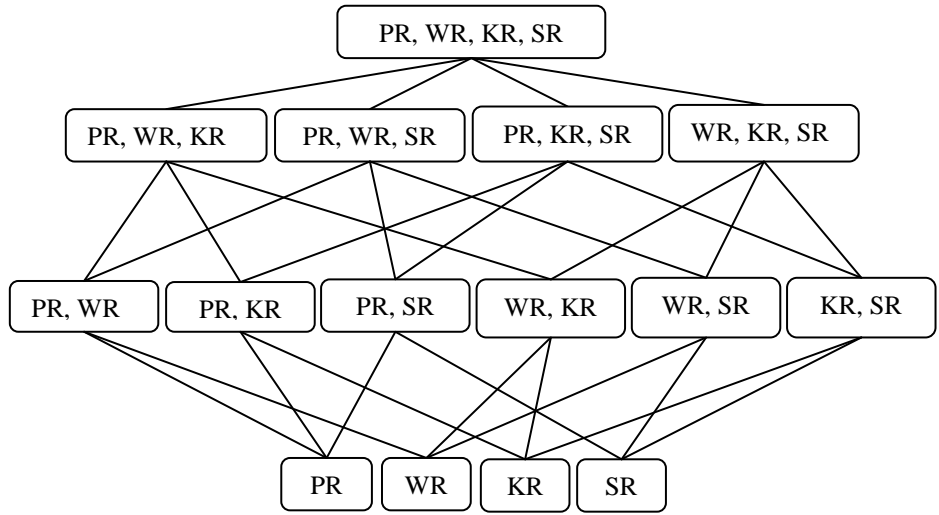

**Fig. 1.** Superiority relationship among rating categories.

In order to understand Fig. 1, the first level (top) suggests that all ratings count equally, whereas the fourth line (bottom), suggests an absolute preference to personal experience (PR), over whitelisted acquaintances (WR), over mere acquaintances (KR), and finally over strangers (SR). Thus, nodes on the left have precedence over nodes on the right. Furthermore, combinations of nodes from different levels can be made, provided that each rating source (PR, WR, KR, SR) is included only once. For example, one can combine node {PR, WR} from the third level with nodes {KR}, {SR} from the bottom level. This means that personal experience and experience of absolutely trusted acquaintances is treated equally, and both of them are preferred over ratings from mere acquaintances and over ratings from strangers.

As soon as the requesting agent A collects all the available ratings, it has to decide upon which of them will participate in the estimation. In order to do this, it has first to indicate which of them are eligible for participating in the final reputation value of agent X; namely a combination of four coefficients: $R_X = \{PR_X, WR_X, KR_X, SR_X\}$. Hence, DISARM, as already discussed, uses confidence and transaction value in order to help agents to discard the collected ratings. Besides, it is important to take into account ratings that were made by confident trusters, since their ratings are more likely to be right. Additionally, confident trusters, that were interacting in an important for them transaction, are even more likely to report truthful ratings. This assumption led to the following defeasible rules that define which ratings will be eligible for the reputation estimation and which not, according to the confidence and the transaction values, yet confidence and importance values are not involved in the estimation itself.

$r_{26}$:   *eligible_rating(rating→?id$_x$, truster→?a, trustee→ ?x) :=*

  *confidence_threshold(?conf), transaction_value_threshold(?tran),*

  *rating(id→?id$_x$, truster→?a, trustee→ ?x,*

    *confidence→?conf$_x$, transaction_value→?tran$_x$),*

  *?conf$_x$ >= ?conf, ?tran$_x$ >= ?tran.*

$r_{27}$:   *eligible_rating(rating→?id$_x$, truster→?a, trustee→ ?x) :=*

  *confidence_threshold(?conf), transaction_value_threshold(?tran),*

  *rating(id→?id$_x$, truster→?a, trustee→ ?x,*

    *confidence→?conf$_x$, transaction_value→?tran$_x$),*

  *?conf$_x$ >= ?conf.*

$r_{28}$:   *eligible_rating(rating→?id$_x$, truster→?a, trustee→ ?x) :=*

  *confidence_threshold(?conf), transaction_value_threshold(?tran),*

  *rating(id→?id$_x$, truster→?a, trustee→ ?x,*

    *confidence→?conf$_x$, transaction_value→?tran$_x$),*

  *?tran$_x$ >= ?tran.*

*$r_{26}$>$r_{27}$>$r_{28}$*

To this end, rule $r_{26}$ indicates that if both the truster's confidence and transaction importance are high, according to the user's threshold, then that rating will be eligible for the estimation process. Rule $r_{27}$, on the other hand, indicates that even if the transaction value is lower than the threshold, it doesn't matter so much if the truster's confidence is high. Rule $r_{28}$, finally, indicates that if there are only ratings with high transaction value then they should be eligible. In any other case, the rating should be omitted. Notice that the above rules are defeasible and they all conclude positive literals. However, these literals are conflicting each other, for the same pair of agents (truster and trustee), since we want in the presence e.g. of personal experience to omit strangers' ratings. That's why there is also a superiority relationship between the rules. The conflict set is formally determined as follows:

$C[eligible\_rating(truster \rightarrow ?a, trustee \rightarrow ?x)] =$

$\{ \neg eligible\_rating(truster \rightarrow ?a, trustee \rightarrow ?x) \} \cup$

$\{ eligible\_rating(truster \rightarrow ?a_1, trustee \rightarrow ?x_1) \mid ?a \neq ?a_1 \wedge ?x \neq ?x_1 \}$

Moreover, even if it is defined which ratings are eligible, the final choice is up to the requesting agent A's personal strategy. The criterion for this final choice, as already mentioned, is time. Other agents will prefer to take into account all eligible ratings whereas others will move one step further indicating which of the eligible ratings, e.g. the newest, will finally participate in the estimation. For instance, rules $r_{29}$, $r_{29'}$ and $r_{29''}$ are examples of such a decision; $r_{29}$ indicates that given a time period (from date-time to date-time) then only rating in this interval will count, $r_{29'}$ indicates that only the latest (from a specific time point onwards) will count whereas $r_{29''}$ indicates that only ratings reported back to a time window will count (where now() returns the current time point).

$r_{29}$:     $count\_rating(rating \rightarrow ?id_x, truster \rightarrow ?a, trustee \rightarrow ?x) :=$

         $time\_from\_threshold(?ftime), time\_to\_threshold(?ttime),$

         $rating(id \rightarrow ?id_x, t \rightarrow ?t_x, truster \rightarrow ?a, trustee \rightarrow ?x),$

         $?ftime <= ?t_x <= ?ttime.$

$r_{29}'$:  count_rating(rating→?$id_x$, truster→?a, trustee→ ?x) :=

    time_from_threshold(?ftime),

    rating(id→?$id_x$, t→?$t_x$, truster→?a, trustee→ ?x),

    ?ftime <=?$t_x$.

$r_{29}''$:  count_rating(rating→?$id_x$, truster→?a, trustee→ ?x) :=

    time_window(?wtime),

    rating(id→?$id_x$, t→?$t_x$, truster→?a, trustee→ ?x),

    now() - ?wtime <=?$t_x$.

DISARM, taking all the above into account, eventually categorizes the ratings (rules $r_{30}$ to $r_{33}$) into the previously defined categories and takes the final decision about which of the ratings can actually participate in the estimation process for the final reputation value $R_X$ (rules $r_{34}$ to $r_{36}$).

$r_{30}$:  count_pr (rating→?id, trustee→?x) :-

    eligible_rating(rating → ?id, truster→?self, trustee→ ?x),

    count_rating(rating → ?id, truster→?self, trustee→ ?x).

$r_{31}$:  count_wr (rating →?id, trustee→?x) :-

    known(agent→?k), WL (trustee →?k),

    eligible_rating(rating → ?id, truster→?k, trustee→ ?x),

    count_rating(rating→?id, truster→?k, trustee→ ?x).

$r_{32}$:  count_kr (rating →?id, trustee→?x) :-

    known(agent→?k),

    not(BL (trustee →?k)), not(WL (trustee →?k)),

    eligible_rating(rating→?id, truster→?k, trustee→ ?x),

    count_rating(rating→?id, truster→?k, trustee→ ?x).

$r_{33}$:  count_sr (trustee→?x, rating→?id) :-

$$eligible\_rating(rating \rightarrow ?id, truster \rightarrow ?s, trustee \rightarrow ?x),$$

$$count\_rating(rating \rightarrow ?id, truster \rightarrow ?s, trustee \rightarrow ?x),$$

$$not(known(agent \rightarrow ?s)).$$

Rules $r_{30}$ to $r_{33}$, categorize the counted ratings in $PR_X$ (direct experience), $WR_X$ (known and trusted / white-listed witness), $KR_X$ (just known witness) and $SR_X$ (strangers' witness), respectively. In $r_{32}$ and $r_{33}$, we use negation as failure. Specifically, in $r_{32}$ if an agent ?k cannot be found both in the whitelist and the blacklist, it is considered as a known witness. Furthermore, in $r_{33}$ if *known*() fails during execution then not(*known*()) will succeed, in order to determine which agents are considered totally strangers. Notice that the above rules are strict ones, i.e. their conclusions cannot be disputed.

The final decision making process for the $R_X$ is based on a relationship theory among the rating categories. In Fig. 1, we presented the complete relationship among all rating categories, whereas, below, we present three potential theories based on that relationship. In the first theory, all categories count, hence, if ratings from all of them are available ($r_{34}$ to $r_{37}$), then they will all participate in the final reputation estimation. To this end, if one of them is missing, then the other two are combined, whereas if just one category is available, then just that will be taken into account. This theory is equivalent to the first row in Fig. 1, namely the combination {PR, WR, KR, SR}.

$r_{34}$:     *participate*(*trustee*→?x, *rating*→?id) :=

           *count_pr*(*trustee*→?x, *rating*→ ?id).

$r_{35}$:     *participate*(*trustee*→?x, *rating*→?id) :=

           *count_wr*(*trustee*→?x, *rating*→ ?id).

$r_{36}$:     *participate*(*trustee*→?x, *rating*→?id) :=

           *count_kr*(*trustee*→?x, *rating*→ ?id).

$r_{37}$:      *participate(trustee→?x, rating→?id)* :=

         *count_sr(trustee→?x, rating→ ?id)*.

In the rest two theories, opinions from different categories conflict each other (conflicting literals), therefore the conflict is being resolved via adding superiority relationships. Specifically, personal opinion is the most important, and then comes white-listed agents' opinion, then simply known agents' and then strangers'. We will present only the superiority relationships and we will not duplicate the rules. The conflict set (for both theories) is:

*C[participate(trustee→?x)]* =

         { ¬ *participate(trustee→?x)* } ∪

         { *participate(trustee→?x$_1$)* |*?x ≠ ?x$_1$* }

In the second theory, the priority relationship among the rules is based on the fact that an agent relies on its own experience if it believes it is sufficient, if not it acquires the opinions of others, much as do humans in real life. This theory is equivalent to the last row in Fig. 1, namely the combination {PR}, {WR}, {KR}, {SR}.

*$r_{34}$>$r_{35}$>$r_{36}$>$r_{37}$*

In the third theory, on the other hand, if direct experience is available (PR), then it is preferred to be combined with ratings from well trusted agents (WR). On the other hand, if personal experience is not available, then ratings from well trusted agents is preferred over just known agents, which is preferred over ratings coming from strangers. In the end, if nothing of the above is available, DISARM acts as a pure witness system. This theory is equivalent to the combination of the first node of the third row in Fig. 1, with the two last nodes of the last row, namely the combination {PR, WR}, {KR}, {SR}.

*$r_{34}$>$r_{37}$, $r_{34}$>$r_{36}$, $r_{35}$>$r_{36}$, $r_{35}$>$r_{37}$, $r_{36}$>$r_{37}$*

### 3.3.4 Estimating Reputation

Agent A eventually reaches on a decision upon which rating is going to participate in the estimation ($R_X$ = {$PR_X$, $WR_X$, $KR_X$, $SR_X$}), according to the chosen relationship theory, as discussed above. In this context, in order to cross out outliers, extremely positive or extremely negative values, the rating values are logarithmically transformed. Outliers are rating values that differ significantly from the mean (a central tendency) and, thus, they can have a large impact on the estimation process that could mislead agents. To this end, the most important feature of the logarithm is that, relatively, it moves big values closer together while it moves small values farther apart and, thus, rating data are better analyzed. More specifically, many statistical techniques work better with data that are single-peaked and symmetric while it is easier to describe the relationship between variables when it is approximately linear. Thus, when these conditions are not true in the original data, they can often be achieved by applying a logarithmic transformation.

To this end, each rating is normalized ($r \in [-1,1]$ | $-1 \equiv$ terrible, $1 \equiv$ perfect), by using 10 as the logarithm base. Thus, the final reputation value ranges from -1 to +1, where -1, +1, 0 stand for absolutely negative, absolutely positive and neutral, respectively, which means that an agent's reputation could be either negative or positive. Hence, the final reputation value $R_X$ is a function $\Im$ that combines the transformed ratings for each available category:

$$R_X = \Im\left(PR_x, WR_x, KR_x, SR_x\right) \quad (1)$$

Moreover, since DISARM aims at simulating human behavior, it allows agents to determine what and how important is each rating parameter for them. In other words, an agent may consider validity more important than all, while it may not care at all about the outcome feeling of the interaction. An example could be the following: {*response time→20%,*

*validity*→*50%*, *completeness*→*10%*, *correctness*→*10%*, *cooperation*→*10%*, *outcome feeling*→*0%*}. Hence, agents are allowed to provide specific weights ($w_i$, $i \in [1, 6]$) that will indicate their personal preferences according the ratings' coefficients. Formula 2, which is the modified Formula 1, calculates the weighted normalized values:

$$R_X = \Im\left[\frac{AVG\left(w_i \times \log\left(pr_X^{coefficient}\right)\right)}{\sum_{i=1}^{6} w_i}, \frac{AVG\left(w_i \times \log\left(wr_X^{coefficient}\right)\right)}{\sum_{i=1}^{6} w_i}, \frac{AVG\left(w_i \times \log\left(kr_X^{coefficient}\right)\right)}{\sum_{i=1}^{6} w_i}, \frac{AVG\left(w_i \times \log\left(sr_X^{coefficient}\right)\right)}{\sum_{i=1}^{6} w_i}\right],$$

$coefficient = \{response\_time, validity, completeness, correctness, cooperation, outcome\_feeling\}$ (2)

Moving one step further, we try to understand deeper the relationship among the rating categories that participate in the estimation. It is up to the chosen relationship theory, presented in the previous subsection, which categories will participate, yet there is no clue about their percentage use in the estimation. To this end, in DISARM the user, through his/her agent A, is able to set what we call the "social trust weights" ($\pi_p$, $\pi_w$, $\pi_k$, $\pi_s$). These weights specify the balance between personal experience ($\pi_p$) and witness reputation ($\pi_w$, $\pi_k$, $\pi_s$). Hence, the final reputation value $R_X$ is calculated according to which experience is more important for the end user (Formula 3).

$$R_X = \Im\left[\begin{array}{c} \frac{\pi_p}{\pi_p + \pi_w + \pi_k + \pi_s} \times \frac{AVG\left(w_i \times \log\left(pr_X^{coefficient}\right)\right)}{\sum_{i=1}^{6} w_i}, \frac{\pi_w}{\pi_p + \pi_w + \pi_k + \pi_s} \times \frac{AVG\left(w_i \times \log\left(wr_X^{coefficient}\right)\right)}{\sum_{i=1}^{6} w_i}, \\ \frac{\pi_k}{\pi_p + \pi_w + \pi_k + \pi_s} \times \frac{AVG\left(w_i \times \log\left(kr_X^{coefficient}\right)\right)}{\sum_{i=1}^{6} w_i}, \frac{\pi_s}{\pi_p + \pi_w + \pi_k + \pi_s} \times \frac{AVG\left(w_i \times \log\left(sr_X^{coefficient}\right)\right)}{\sum_{i=1}^{6} w_i} \end{array}\right],$$

$coefficient = \{response\_time, validity, completeness, correctness, cooperation, outcome\_feeling\}$ (3)

Finally, since time is an important aspect in reputation, DISARM allows time not only to be used for discarding available ratings but also to be used in the estimation itself. It is generally accepted that more recent ratings "weigh" more since they represent the latest activity of a specific agent. In order to include this aspect in the final reputation value, each rating at time t ($t < t_{now}$) is multiplied with t itself, as shown below. So, time becomes a sort of weight; the larger

(i.e. the most recent), the more it weighs. Hence, Formula 4 represents DISARM's final metric. Moreover, mention that a potential example of this formula, the simplest one for function $\Im$, could be the summation; in the sense that all categories participate additively in the final value, each one with its own weight.

$$R_X = \Im \begin{bmatrix} \dfrac{\pi_p}{\pi_p+\pi_w+\pi_k+\pi_s} \times \dfrac{AVG\left(w_i \times \dfrac{\sum_{\forall t<t_{current}}[\log(pr_X^{coefficient}(t))\times t]}{\sum_{\forall t<t_{now}}t}\right)}{\sum_{i=1}^{6}w_i}, \\ \dfrac{\pi_w}{\pi_p+\pi_w+\pi_k+\pi_s} \times \dfrac{AVG\left(w_i \times \dfrac{\sum_{\forall t<t_{current}}[\log(wr_X^{coefficient}(t))\times t]}{\sum_{\forall t<t_{now}}t}\right)}{\sum_{i=1}^{6}w_i}, \\ \dfrac{\pi_k}{\pi_p+\pi_w+\pi_k+\pi_s} \times \dfrac{AVG\left(w_i \times \dfrac{\sum_{\forall t<t_{current}}[\log(kr_X^{coefficient}(t))\times t]}{\sum_{\forall t<t_{now}}t}\right)}{\sum_{i=1}^{6}w_i}, \\ \dfrac{\pi_s}{\pi_p+\pi_w+\pi_k+\pi_s} \times \dfrac{AVG\left(w_i \times \dfrac{\sum_{\forall t<t_{current}}[\log(sr_X^{coefficient}(t))\times t]}{\sum_{\forall t<t_{now}}t}\right)}{\sum_{i=1}^{6}w_i} \end{bmatrix},$$

$coefficient = \{response\_time, validity, completeness, correctness, cooperation, outcome\_feeling\}$ (4)

### 3.3.5 Measuring estimation confidence

As already mentioned, DISARM also studies the variability of the ratings that were finally taken into account as a measure about the confidence of the estimation itself. For this purpose, we use standard deviation. It measures the amount of variation or dispersion from the average. Yet, in addition to expressing the variability of a population, the standard deviation is commonly used to measure confidence in statistical conclusions. In other words, the standard deviation is a measure of how spread out numbers are. A low standard deviation indicates that the data points (here ratings) tend to be very close to the mean, the expected value, hence it is more likely the estimation to be closer to the agent's actual behavior. On the other hand, a high standard deviation indicates that the data points (ratings) are spread out over a large range of values and, thus, it is difficult to predict the agent's behavior. Formula 5 presents the standard deviation metric used in DISARM, where N represents the total amount of used (in formula 4) ratings (r).

$$\sigma = \sqrt{\frac{1}{N}\sum_{j=1}^{N}(r_X^{coefficient} - \mu(r_X^{coefficient}))^2}, r_X^{coefficient} \in pr_X^{coefficient} \cup wr_X^{coefficient} \cup kr_X^{coefficient} \cup sr_X^{coefficient},$$

$$coefficient = \{response\_time, validity, completeness, correctness, cooperation, outcome\_feeling\} \quad (5)$$

More specifically, in DISARM the above formula does not participate in the estimation process itself nor affect, in any way, the reputation value. Its role is to act complement to DISARM's final metric (formula 4), in order to indicate the probability the estimated value (formula 4) to be close to reality. The motivation behind the use of the standard deviation formula (formula 5) was the fact that although reputation models provide an estimated reputation value they do not provide any clue about how likely is this estimation to reflect the agent's true behavior. To this end, DISARM provides an additional tool (formula 5) in order to assist agents, and thus their users, to make the best for them choices.

## 4 Evaluation

For evaluation purposes, regarding DISARM, we combined two testbed environments, adopted from [21, 20], previously used in [27], and [24]. These testbeds are quite similar, just with slight differences in number of participants and simulation rounds. In this context, we preserved the testbed design but slightly changed the evaluation settings, taking into account the data provided in previous works. Below, a description of the testbed is given, and next the methodology and the experimental settings for our experiments are also presented. The testbed environment is a multi-agent system consisting of agents providing services and agents that use these services. We assume that the performance of a provider, and effectively its trustworthiness, is independent from the service that is provided. In order to reduce the complexity of the testbed's environment, it is assumed that there is only one type of service in the testbed and, as a result, all the providers offer the same service.

Nevertheless, the performance of the providers, such as the quality of the service in terms of satisfaction, response time, etc., differs and determines the utility that a consumer gains from each interaction (called UG≡utility gain). The value of UG varies from 0 to 10 and it depends on the level of performance of the provider in that interaction. A provider agent can serve many users at a time. After an interaction, the consumer agent rates the service of the provider based on the level of performance and the quality of the service it received. It is assumed that all agents exchange their information honestly in this testbed. This means an agent (as a witness) provides its true ratings as they are without any modification. Each agent interaction is a simulation round. Events that take place in the same round are considered simultaneous and, thus, the round number is used as the time stamp for events and ratings.

In this context, for implementation purposes, we use EMERALD [28], a framework for interoperating knowledge-based intelligent agents in the Semantic Web. This framework is built on top of JADE [4], a reliable and widely used multi-agent framework. EMERALD was chosen since it provides a safe, generic, and reusable framework for modeling and monitoring agent communication and agreements. Moreover, it proposes, among others, the use of Reasoners [29]. Reasoners are agents that offer reasoning services to the rest of the agent community. A Reasoner can launch an associated reasoning engine, in order to perform inference and provide results. EMERALD supports a number of Reasoners but most important for the purposes of this article are the four Reasoners that use defeasible reasoning; among them is the DR-Reasoner (based on DR-Device defeasible logic system [3]), the defeasible Reasoner that was used for the evaluation. Additionally, EMERALD provides an advanced yellow paper service, called AYPS, that is responsible for recording and representing information related to registered in the environment agents, namely their name, type, registration time and activity. This information is dynamically stored in the AYPS agent's database. Hence, the service is able to retrieve up-to-date information at any time.

Hence, even if DISARM, or any other distributed model, is a distributed reputation model, agents that use it are able to send requests to AYPS in order to get first a list of potential partners. Next, they will use the DISARM model in order to estimate reputation for one or more of them in order to find the most appropriate partner. Of course, it is not necessary to use such services; it is up to each agent's personal strategy how it will locate potential partners. The more an agent knows the environment, the better it can choose providers and, thus, the more utility gains. In this context, agents in the environment are free to ask others for their opinion (ratings), hence each agent requests the service from the most trustworthy and reliable provider according to it. Furthermore, concerning DISARM's final metric (formula 4), in this section we adopt the addition as shown below:

$$R_X = \frac{\pi_p}{\pi_p+\pi_w+\pi_k+\pi_s} \times \frac{AVG\left(w_i \times \frac{\sum_{\forall t<t_{current}}[\log(pr_X^{coefficient}(t))\times t]}{\sum_{\forall t<t_{now}} t}\right)}{\sum_{i=1}^{6} w_i} + \frac{\pi_w}{\pi_p+\pi_w+\pi_k+\pi_s} \times \frac{AVG\left(w_i \times \frac{\sum_{\forall t<t_{current}}[\log(wr_X^{coefficient}(t))\times t]}{\sum_{\forall t<t_{now}} t}\right)}{\sum_{i=1}^{6} w_i} +$$

$$\frac{\pi_k}{\pi_p+\pi_w+\pi_k+\pi_s} \times \frac{AVG\left(w_i \times \frac{\sum_{\forall t<t_{current}}[\log(kr_X^{coefficient}(t))\times t]}{\sum_{\forall t<t_{now}} t}\right)}{\sum_{i=1}^{6} w_i} + \frac{\pi_s}{\pi_p+\pi_w+\pi_k+\pi_s} \times \frac{AVG\left(w_i \times \frac{\sum_{\forall t<t_{current}}[\log(sr_X^{coefficient}(t))\times t]}{\sum_{\forall t<t_{now}} t}\right)}{\sum_{i=1}^{6} w_i},$$

$coefficient = \{response\_time, validity, completeness, correctness, cooperation, outcome\_feeling\}$ (4)

To this end, taking all the above into account, the testbed in each experiment is populated with provider and consumer agents. Each consumer agent is equipped with a particular trust model (a centralized approach is also included), which helps it select a provider when it needs to use a service. The only difference among consumer agents is the trust models that they use, so the utility gained by each agent through simulations will reflect the performance of its trust model in selecting reliable providers for interactions. As a result, the testbed records the UG of each interaction with each trust model used. Consumer agents without the ability to choose a trust model will randomly select a provider from the list. Furthermore, in order to obtain an accurate result for performance comparisons between trust models, each one will be employed by a large but equal number of consumer agents.

**Table 1.** Testbed environment.

|  | *Density in the environment* |  | *Density in the environment* |
|---|---|---|---|
| *Service Providers* |  | *Service Consumers* |  |
| Good providers | 15% | DISARM | 14.28% |
| Ordinary providers | 30% | Social Regret | 14.28% |
| Intermittent providers | 15% | Certified Reputation | 14.28% |
| Bad providers | 40% | CRM | 14.28% |
|  |  | FIRE | 14.28% |
|  |  | HARM | 14.28% |
|  |  | NONE | 14.28% |

Hence, Table 1 displays the testbed environment; the four types of service providers used, namely good, ordinary, bad and fickle (or intermittent). The first three provide services according to the assigned mean value of quality with a small range of deviation. Fickle agents, on the other hand, cover all possible outcomes randomly. Finally, in this section, we compare DISARM, Social Regret, Certified Reputation, CRM, FIRE, HARM [27] and NONE (no trust mechanism, randomly selected providers). We used HARM although it is a centralized approach since it is a rule-based model using temporal defeasible reasoning. In this context, taking into account all the available data, Fig 2 depicts the overall ranking regarding the utility gained for all models, even for absence of model, namely NONE (random selection).

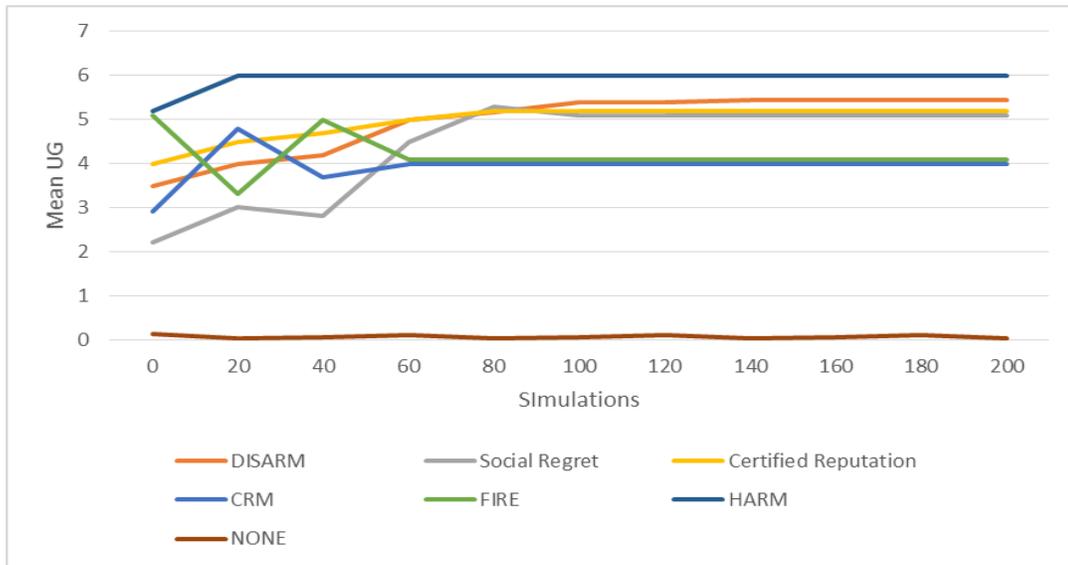

**Fig. 2.** Mean Utility Gained Ranking.

As shown in Fig. 2 NONE performance is poor and, as expected, consistently the lowest. HARM, on the other hand, is consistently the highest. This is not surprising since HARM is a rule-based, centralized model. Hence, it is able to gather ratings about all interactions in the system as opposed to the rest distributed models, where locating rating is a challenging task by itself. This allows agents using HARM to achieve higher performance right from the first interactions. Concerning, the distributed models, it is clear that DISARM, Certified Reputation and Social Regret gain a quite high UG value, yet they unable to reach the performance of centralized models like HARM. Among distributed models, DISARM achieves a slight high performance, mainly due to the fact that it is using a dynamic (defeasible) reputation estimation mechanism that enables agents to take more intuitive decisions and, thus, increase their performance. Furthermore, DISARM enables agents to get familiarized with the environment faster as opposed to CRM and FIRE which, as shown in Fig. 2, need more time to know the environment and stabilize their performance.

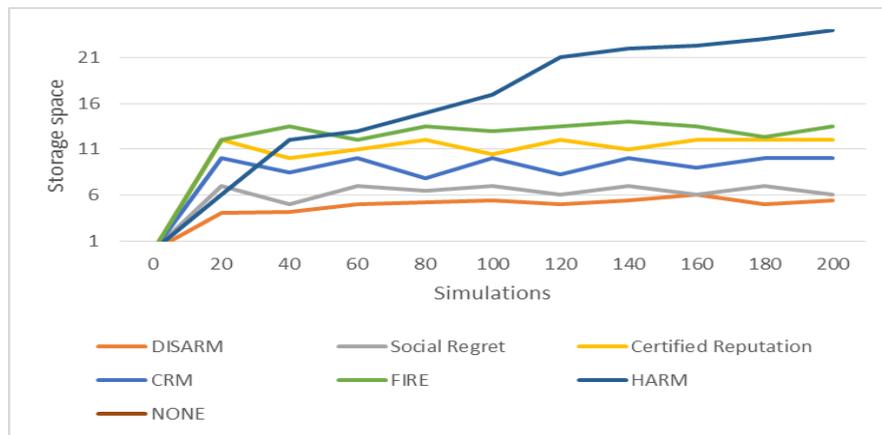

**Fig. 3.** Storage space grow.

Although, centralized models achieve higher UG score they have significant limitations in terms of execution time and storage space. This is not surprising since centralized models are, usually, managed by a single agent. This manager has to store all ratings in the system, which are increased over time, and to respond to an increasing number of requests, leading to bottle-neck effect. On the other hand, distributed models store just their own ratings and those obtained by witnesses which are far less than the whole available ratings in the system. Additionally, these extra witness ratings could be erased after use, releasing space. As shown in Fig. 3, HARM, being a centralized model, needs much more space than distributed models, reaching even the double. On the other hand, models like DISARM and Social Regret that take into account social aspects, need less space, even from other distributed approaches, since they collect less ratings. Hence, the more ratings used by a model the more space (and usually time) is needed.

## 5   Related Work

Trust and reputation represent a significant aspect in modern multi-agent systems. An interesting and very challenging active research area is already focused on them; various models and metrics have already been proposed in order to deal with the challenging decision making processes in the agent community [17, 15, 37]. Reputation is used to build trust among agents, minimizing the risk involved in the transactions and increasing users' confidence and satisfaction. Hence, since the best decisions are those that taken under the minimum risk, trust and reputation models support agents to take promising decisions regarding potential partners.

To this end, one of the first, if not the first, model that used the idea of witness reputation was a decentralized trust model, called Regret [47, 45]. Regret is, actually, one of the most representative trust and reputation models in multi-agent systems. It combines witness reports and direct interaction experience in order to provide reputation values. Additionally, in Regret ratings are dealt with respect to time; old ratings are given less importance compared to new ones. An evolution of Regret, a primary attempt to locate witnesses' ratings, called Social Regret [46], was also presented by the authors. Social Regret is a reputation system oriented to e-commerce environments that incorporates the notion of social graph. More specifically, Social Regret groups agents with frequent interactions among them and considers each one of these groups as a single source of reputation values. In this context, only the most representative agent within each group is asked for information. To this end, a heuristic is used in order to find groups and to select the best agent to ask.

Social Regret, similarly to DISARM, is one of the rare cases that the social dimension of agents is taken into account. Yet, Social Regret does not reflect the actual social relations among agents, like DISARM, but rather attempts to heuristically reduce the number of queries to be done in order to locate ratings. Taking into account the opinion of only one agent of each group is a severe disadvantage since the most agents are marginalized, distorting reality. However, both Regret and DISARM recognize the importance of time and take into account both personal and witness ratings. Yet, only DISARM allows agents to decide on their own about what they consider important regarding time. Additionally, only DISARM provides a knowledge-based mechanism, promoting a nonmonotonic, more flexible, human-like approach.

Another popular distributed model is FIRE [21]. FIRE integrates four types of trust and reputation, namely interaction trust, role-based trust, witness reputation and certified reputation. Interaction trust and witness reputation are, as in DISARM, an agent's past experience from direct interactions and reports provided by witnesses about an agent's behavior, respectively. Role-based trust, on the other hand, is trust defined by various role-based relationships between the agents whereas certified reputation is third-party references provided by the target agents. The aforementioned values are combined into a single measure by using the weighted mean method. FIRE similar to DISARM recognizes the need for hybrid models that will take into account more than one source for the final reputation estimation. Yet, although FIRE take into account more sources than DISARM, it uses a weak computation model for the final combination and reputation estimation. DISARM, on the other hand, provides a human-like knowledge-based mechanism, based on defeasible logic that let agents take into account the most promising available rating in order to predict the future behavior of a potential partner. Additionally, only DISARM takes in account the social dimension of multi-agent systems with respect to time.

Another remarkable reputation model is Certified Reputation [20], a decentralized reputation model, like DISARM, involving each agent keeping a set of references given to it from other agents. In this model, each agent is asked to give certified ratings of its performance after every transaction. The agent then chooses the highest ratings and stores them as references. Any other agent can then ask for the stored references and calculate the agent's certified reputation. This model overcomes the problem of initial reliability in a similar way with DISARM. However, opposed to our approach, this model is designed to determine the access rights of agents, rather than to determine their expected performance. Furthermore, it is a witness-based model, whereas DISARM combines both witnesses and direct experience, providing a rule-based methodology to deal with the discrimination issue. Furthermore, although in Certified Reputation agents are freed from the various costs involved in locating witness reports, such as resource, time and communication costs, ratings might be misquoted since it is each agent's responsibility to provide ratings about itself and in the fact the best ones.

TRR (Trust–Reliability–Reputation) trust model [44] allows a software agent to represent both the reliability and the reputation of another agent, merging finally these measures into a global trust evaluation. It uses a dynamically computed weight that represents how an agent considers important the reliability with respect to the reputation when it computes the trust of another agent. Yet, the weight depends on the number of interactions between the two agents, which is actually a problem when these agents have no interaction history. Hence, TRR provides a mechanism for estimating the global reputation value of an agent based on previous interactions. On the other hand, DISARM provides a knowledge-based mechanism that enables each agent to estimate a personalized reputation based on its preferences. Moreover, DISARM takes into account plenty of issues, such as time and social relations, although it does not deal with reliability issues as TRR does. Additionally, DISARM is a nonmonotonic approach based on defeasible logic that provides a more flexible, human-like approach that enables agents not only to estimate a reputation value but also to decide upon their relationships in the community.

CRM (Comprehensive Reputation Model) [24] is another typical distributed reputation model. In CRM the ratings used to assess the trustworthiness of a particular agent can either be obtained from an agent's interaction history or collected from other agents that can provide their suggestions in the form of ratings; namely interaction trust and witness reputation, respectively. CRM is a probabilistic-based model, taking into account the number of interactions between agents, the timely relevance of provided information and the confidence of reporting agents on the provided data. More specifically, CRM, first, takes into account direct interactions among agents, calling the procedure online trust estimation. After a variable interval of time, the actual performance of the evaluated agent is compared against the information provided by other agents in a procedure called off-line. Off-line procedure considers the communicated information to judge the accuracy of the consulting agents in the previous on-line trust assessment process. In other words, in CRM the trust assessment procedure is composed of on-line and off-line evaluation processes. Both CRM and DISARM acknowledge the need for hybrid reputation models taking into account time issues, yet they propose a starkly opposite approach. Additionally, both models use a confidence parameter in order to weight ratings more accurately. However, DISARM takes into account a variety of additional parameters, allowing users to define weights about them. As a result, more accurate and personalized estimations are provided. Furthermore, only DISARM considers the social relations among agents providing a nonmonotonic approach that let them establish and maintain trust relationships, locating quite easily reliable ratings.

Finally, HARM [27], a previous work of us, is a hybrid rule-based reputation model that uses temporal defeasible logic in order to combine interaction trust and witness reputation. Yet, it is a centralized approach which actually overcomes the difficulty to locate witness reports. DISARM, on the other, hand is also a hybrid but distributed model that uses defeasible (yet not temporal) logic in a similar point of view. Actually, DISARM is an updated and extended model based partially on HARM's principles though adapting a decentralized approach. To this end, although both models consider time important, they are dealing with it with a totally different approach. Ratings in HARM are characterized by a time offset property, which indicates the time instances that should pass in order to consider each rating active while each of them counts only for a limited time duration. DISARM uses the time itself in the final estimation formula, letting agents to use a similar to human thinking philosophy that first decides upon which category of rating should be taken into account and then discards ratings included there. Comparing, these two models, we believe that DISARM and HARM, despite their similarities and differences, are two nonmonotonic models that enable agents to improve their effectiveness and intuitiveness in a way more related to the traditional human reasoning for assessing trust in the physical word.

# 6 Conclusions and Future Work

This article presented DISARM, a social, distributed, hybrid, rule-based reputation model which uses defeasible logic. DISARM though appropriate rules, combines interaction trust and witness reputation. Moreover, it limits the common disadvantages of the existing distributed trust approaches, such as locating ratings, by considering the agents acting in the environment as a social network. Hence, each agent is able to propagate its requests in the rest community, locating quite fast ratings from previously known and well-rated agents. Additionally, DISARM's mechanism is based on defeasible logic, modeling the way intelligent agents, like humans, draw reasonable conclusions from inconclusive information, which is one of the main advantages of the model. Actually, it is one of the first models that use nonmonotonic knowledge, in the form of defeasible logic in order to predict agents' future behavior. It is based on well-established estimation parameters [8, 9], such as information correctness, completeness, and validity, agent's response time and cooperation, as well as outcome feeling of the interaction. Hence, DISARM can be adopted in any multi-agent system in the Semantic Web, such as JADE and EMERALD. Finally, we provided an evaluation that illustrates the usability of the proposed model.

As for future directions, first of all, we plan to study further DISARM's performance by comparing it to more reputation models from the literature and use it in real-world applications, combining it also with Semantic Web metadata for trust [10, 11]. Another direction is towards improving DISARM. There are still some open issues and challenges regarding, for instance, rating locating. More technologies could be adopted for these purpose; ontologies, machine learning techniques and user identity recognition and management are some of them.